\documentclass[aps,preprint,preprintnumbers,amsmath,amssymb,floatfix]{revtex4}
\usepackage[german,english]{babel}
\usepackage{bm,amsmath,amsfonts}
\usepackage{amssymb}
\usepackage{graphicx}
\usepackage{xcolor}

\def\beq{\begin{equation}}
\def\eeq{\end{equation}}
\def\beqn{\begin{eqnarray}}
\def\eeqn{\end{eqnarray}}

\begin{document}

\title{How does an interacting many-body system tunnel through a potential barrier to open space?} 

\author{
Axel U. J. Lode$^{1}$\footnote{Email: axel.lode@pci.uni-heidelberg.de},
Alexej I. Streltsov$^{1}$, 
Kaspar Sakmann$^{1}$, \\
Ofir E. Alon$^{2}$,
and Lorenz S. Cederbaum$^{1}$
}

\affiliation{$^1$ Theoretische Chemie, Physikalisch-Chemisches Institut, Universit\"at Heidelberg,\\
Im Neuenheimer Feld 229, D-69120 Heidelberg, Germany}

\affiliation{$^2$ Department of Physics, University of Haifa at Oranim, Tivon 36006, Israel}

\maketitle

\textbf{
The tunneling process in a many-body system is a phenomenon which lies at the very heart of quantum mechanics.
It appears in nature in the form of $\alpha$-decay, fusion and fission in nuclear physics,
photoassociation and photodissociation in biology and chemistry.
A detailed theoretical description of the decay process in these systems is a very cumbersome problem,
either because of very complicated or even unknown interparticle interactions or due to a large number of constitutent particles. 
In this work, we theoretically study the phenomenon of quantum many-body tunneling in a more transparent and controllable physical system, in an ultracold atomic gas.
We analyze a full, numerically exact many-body solution of the Schr\"{o}dinger equation of a one-dimensional system with repulsive interactions tunneling to open space.
We show how the emitted particles dissociate or fragment from the trapped and coherent source of bosons: the overall many-particle decay process is a quantum interference of 
single-particle tunneling processes emerging from sources with \textit{different particle numbers}
taking place \textit{simultaneously}. 
The close relation to atom lasers and ionization processes allows us to unveil the great relevance of many-body correlations between the emitted and trapped fractions of the wavefunction in the respective processes.
}\\

\begin{figure}[]
\vglue -1.5 truecm
\centering
\begin{center}
\includegraphics[width=10.0cm, angle=-90]{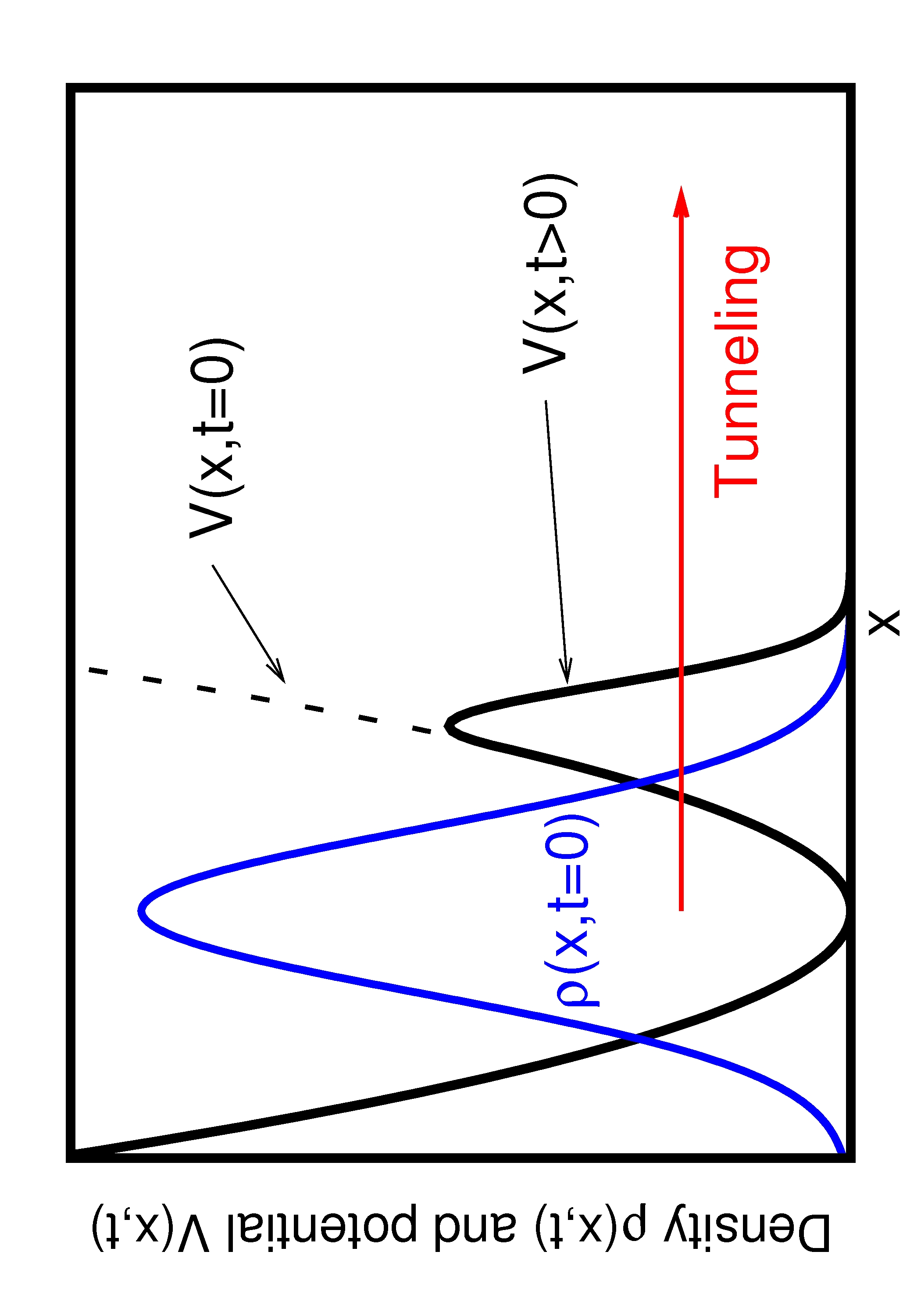}\\
\includegraphics[width=10.0cm, angle=-90]{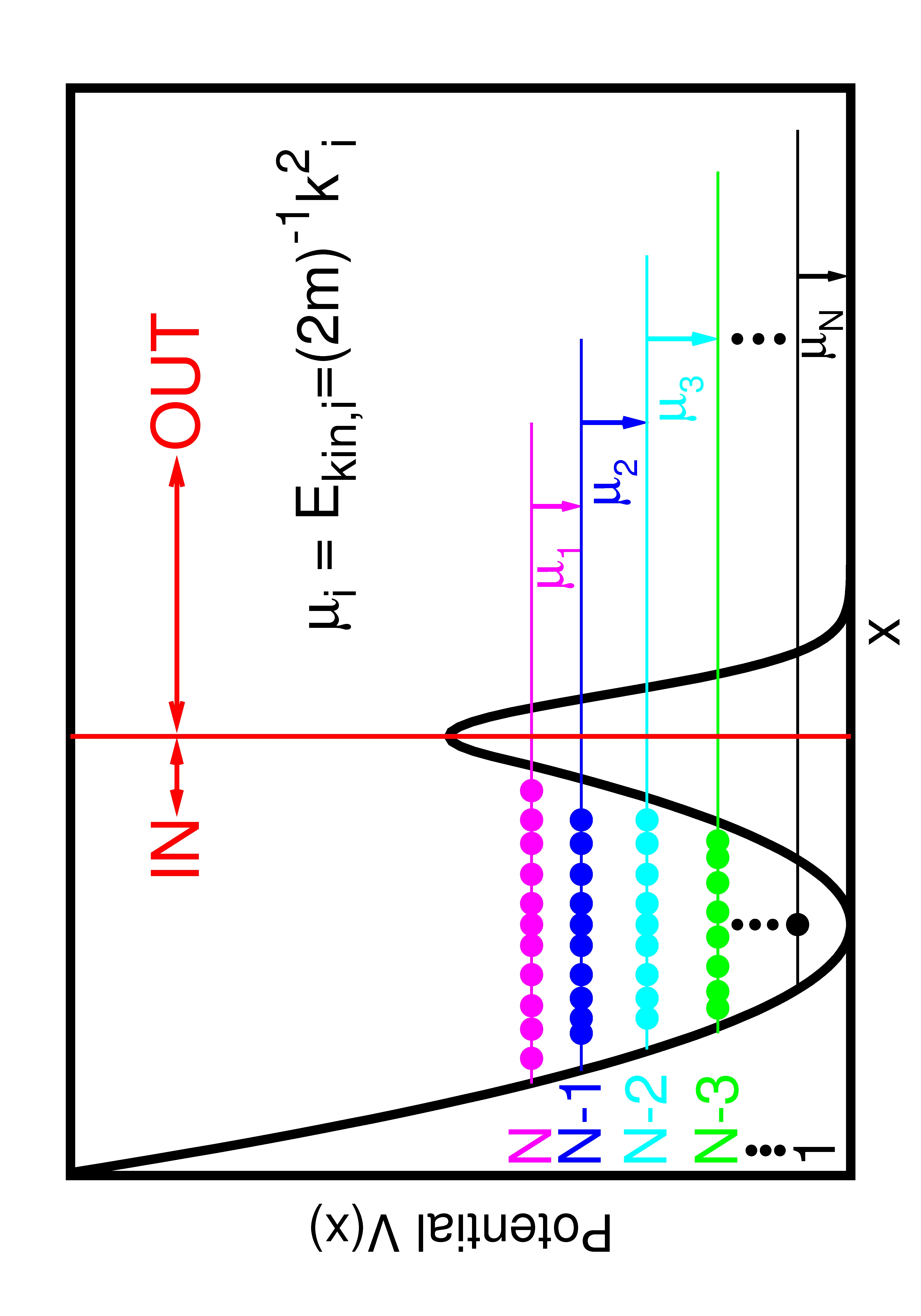}
\end{center}
\caption{{\bf Protocol of the tunneling process. }
\textbf{Top:} A generic density $\rho(x,t<0)$ (blue line) is prepared as the groundstate of a parabolic trap $V(x,t<0)$ (dashed black line). 
The trap is transformed to the open shape $V(x,t\ge0)$ (black line), which allows the system to tunnel to open space. 
\textbf{Bottom:} Sequential mean-field scheme to model the tunneling processes.
The bosons are ejected from ``IN'' to ``OUT'' subspaces (indicated by the red line).  
The chemical potential $\mu_i$ is converted to kinetic energy $E_{kin,i}$. 
\textit{All} the momenta corresponding to the chemical potentials 
$k_i=\sqrt{2mE_{kin,i}}=\sqrt{2m\mu_i}; i=N,N-1,...,1$ appear in the momentum distribution, see Fig.~\ref{FIG2}.
All quantities shown are dimensionless.
}
\label{FIG1}
\end{figure}

The tunneling process has been a matter of discussion \cite{tunstart1,tunstart2,WKB} since the advent of quantum mechanics. In principle, it takes place in all systems whose potential exhibits \textit{classically forbidden}
but \textit{energetically allowed} regions. See, for example, the overview ref.~\cite{tun_book} and Fig.~\ref{FIG1}.
When the potential is unbound in one direction, the quantum nature of the systems
allows them to overcome potential barriers for which they classically would not have sufficient energy
and, as a result, a fraction of the many-particle system is emitted to open space.
For example, in fusion, fission, photoassociation and photodissociation processes, the energetics or life times are of primary interest \cite{alphatun,fistun,fustun,distun,asstun}.
The physical analysis was made under the assumption that the correlation between decay products (i.e., between the remaining and emitted fractions of particles) can be neglected. 
However, it has to be stressed first that, at any finite decay time, the
remaining \textit{and} emitted particles still constitute \textit{one} total many-body wavefunction and, therefore, can be correlated. Second, in contrast to the tunneling of an isolated single particle into open space, which has been amply studied and understood \cite{tun_book}, nearly nothing is known about the tunneling of a many-body system.
In the present study, we demonstrate that the fundamental many-body aspects of quantum tunneling can be studied by monitoring the correlations and coherence in ultracold atomic gases.
For this purpose, the initial state of an ultracold atomic gas of bosons is prepared coherently in a parabolic trapping potential which is subsequently transformed to an open shape allowing for tunneling (see the upper panel of Fig.~\ref{FIG1}). In this tunneling system the correlation between the remaining and emitted particles can be monitored by measuring deviations from the initial coherence of the wavefunction. The close relation to atom lasers and ionization processes allows us to predict coherence properties of atom lasers and to propose the study of ionization processes with tunneling ultracold bosons.

In the past decade, Bose-Einstein condensates (BECs) \cite{RMD1,RMD2} have become a toolbox to study theoretical predictions and
phenomena experimentally with a high level of precision and control \cite{measure}.
BECs are experimentally tunable, the interparticle interactions \cite{feshb}, trap potentials \cite{arb_pot}, number of particles \cite{Raizen_Culling}, their statistics \cite{Selim_statistics} and even dimensionality \cite{1d1,1d2,1d3} are under experimental control.
From the theoretical point of view, the evolution of an ultracold atomic cloud is governed
by the time-dependent many-particle Schr\"{o}dinger equation (TDSE) \cite{TDSE} with a {\it known} Hamiltonian (details in the Methods section).
To study the decay scenario in Fig.~\ref{FIG1}, we solve the TDSE numerically exactly for long propagation times. For this purpose, we use the multiconfigurational time-dependent Hartree method for bosons (MCTDHB) (for details and literature see the Methods section). Our protocol to study the tunneling process of an initially parabolically trapped system into open space is schematically depicted in the upper panel of Fig.~\ref{FIG1}. We restrict our study here to the one-dimensional case, which can be achieved experimentally by adjusting the transverse confinement appropriately. As a first step, we consider $N=2,4$ or $101$ weakly repulsive $^{87}$Rb atoms in the groundstate of a parabolic trap (see the Methods section for a detailed description of the considered experimental parameters).

The initial one-particle density and trap profile are depicted in the upper panel of Fig.~\ref{FIG1}.
Next, the potential is abruptly switched to the open form
$V(x,t\ge0)$ indicated in this figure,
allowing the many-boson system to escape the trap by tunneling through the barrier formed. Eventually, all the bosons escape by tunneling through the barrier, because the potential supports no bound states.

The initial system, i.e., the source of the emitted bosons, is an almost totally coherent state \cite{GP_book}.
The final state decays entirely to open space to the right of the barrier, where the bosons populate many many-body states, related to Lieb-Liniger states \cite{lieb1,lieb2,gaudin},
which are generally not coherent. It is instructive to ask the following guiding questions:
what happens in between these two extremes of complete coherence and complete incoherence?
And how does the correlation (coherence) between the emitted particles and the source evolve?
By finding the answers to these questions we will gain a deeper theoretical understanding of many-body tunneling,
which is of high relevance for future technologies and applied sciences.
In particular, this knowledge will allow us to determine whether the studied ultracold atomic clouds qualify as candidates for atomic lasers \cite{RMD2,atoml1,atoml2,atoml3} or as a toolbox for the study of ionization or decay processes \cite{alphatun,fistun,fustun,distun}.

\begin{figure}[]
    \begin{center}
\includegraphics[angle=-90,width=7.0cm]{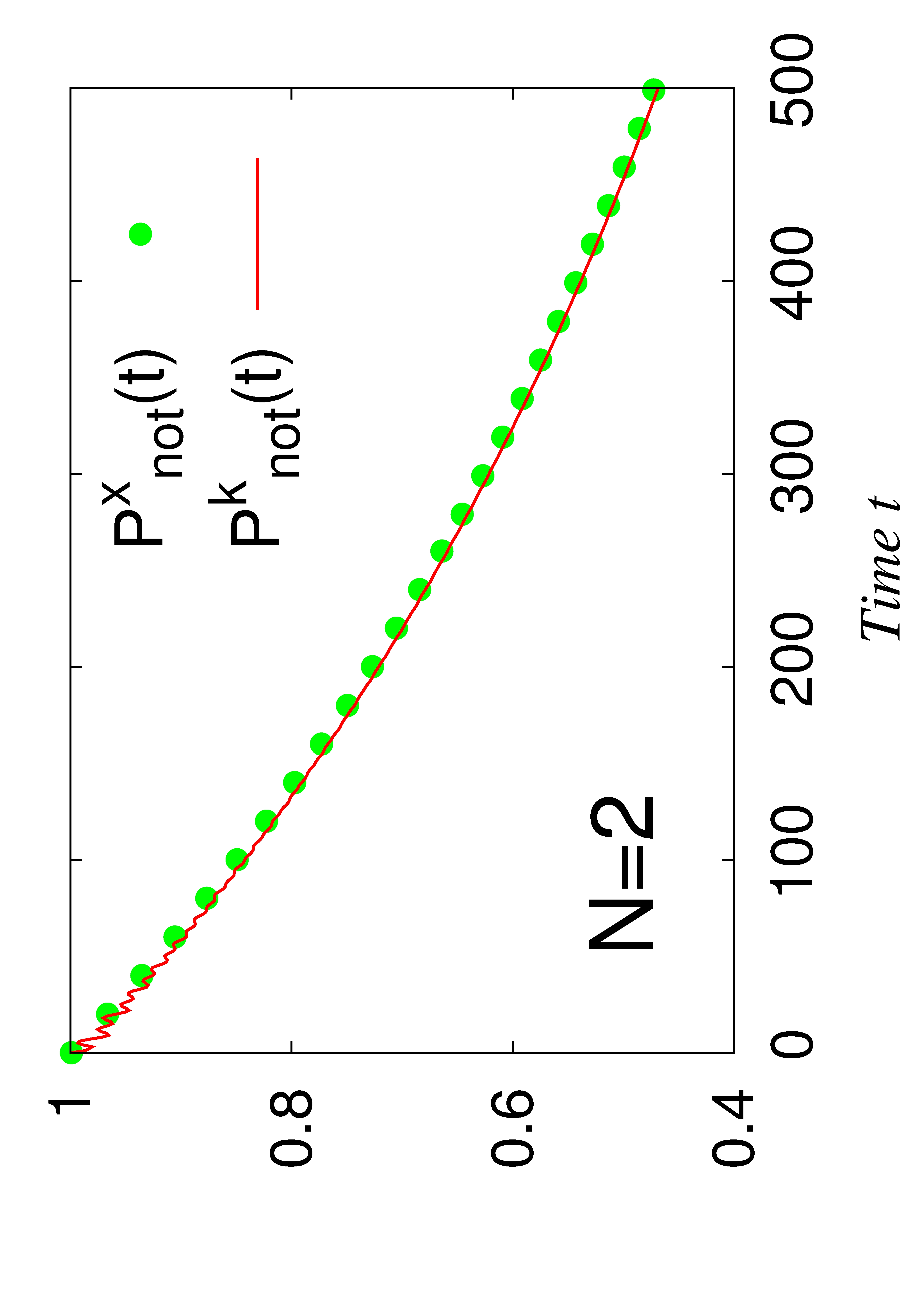}
\includegraphics[angle=-90,width=7.0cm]{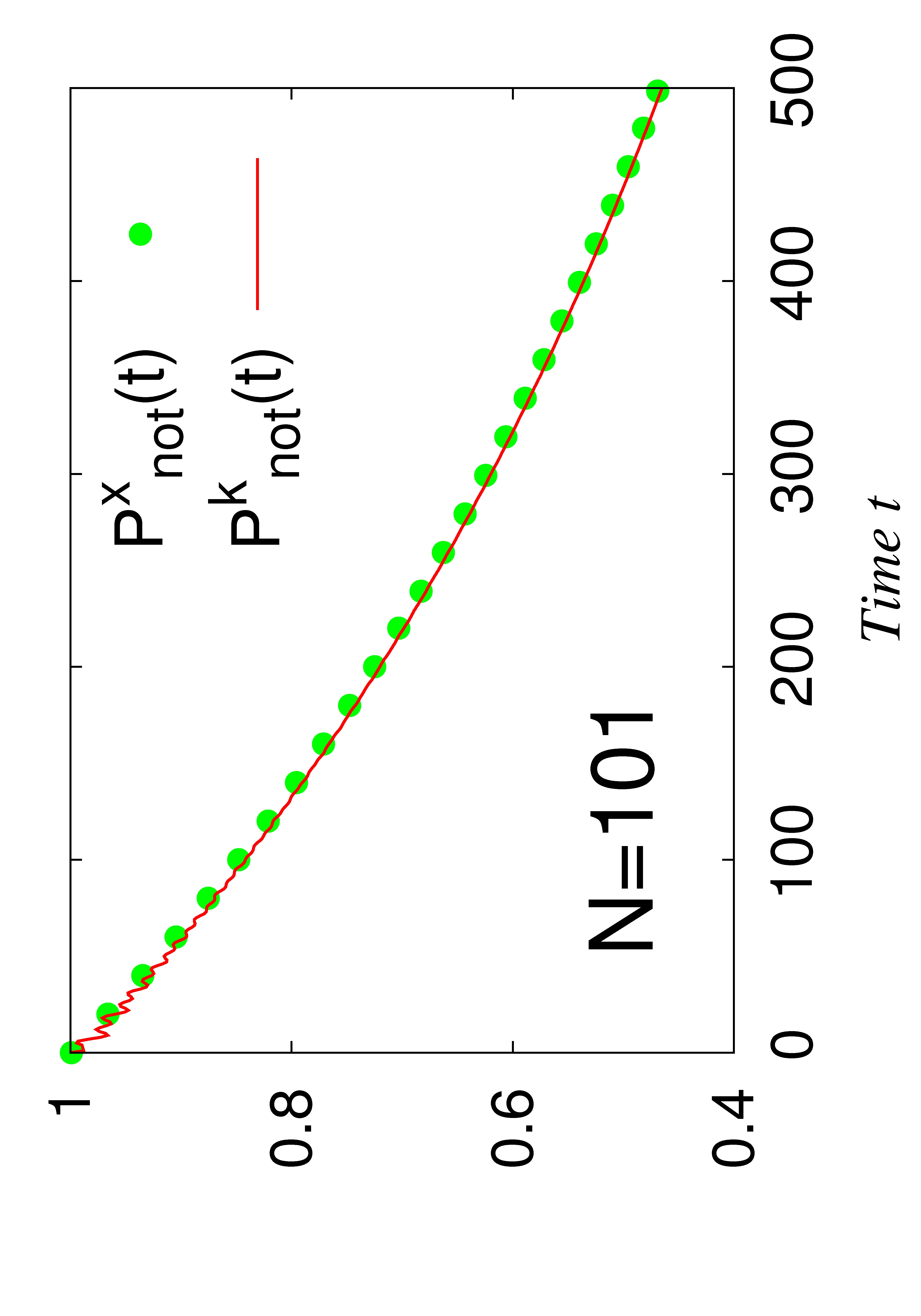}
\end{center}
\caption{{\bf Many-body tunneling to open space is a fundamentally exponential decay process.}
To confirm that the fraction of atoms remaining in the trap decays exponentially with time, we depict the density related nonescape probabilities $P^{x}_{not,\rho}(t)$ in real and $P^{k}_{not,\rho}(t)$ in momentum space, indicated by the respective solid green and red lines.
All quantities shown are dimensionless.
}
    \label{FIG2}
\end{figure}

Since the exact many-body wavefunctions are available at any time in our numerical treatment, 
we can quantify and monitor the evolution of the coherence and correlations of the whole system
as well as between the constituting parts of the evolving wave-packets.
In the further analysis we use the one-particle density in real $\rho(x,t)$ and momentum $\rho(k,t)$ spaces, their natural occupations $\rho^{NO}_i(t)$ and correlation functions $g^{(1)}(k'|k;t)$, see \cite{Glauber,Penrose,RDM_K}.
We defer the details on these quantities to the Methods section.
To study the correlation between the source and emitted bosons we 
decompose the one-dimensional space into
the internal ``IN'' and external ``OUT'' regions with respect to the top of the barrier, as illustrated by the red line and arrows in the lower panel of Fig.~\ref{FIG1}.
This decomposition of the one-dimensional Hilbert space into subspaces allows us
to quantify the tunneling process by measuring the amount $P^x_{not,\rho}(t)$ of particles
remaining in the internal region in real space as a function of time.
In Fig.~\ref{FIG2} we depict the corresponding quantities for $N=2$ and $N=101$ by the green dotted curve.
A first main observation is that the tunneling of bosonic systems to open space
resembles an exponential decay process.

\begin{figure}[]
\vglue -2.0 truecm
    \begin{center}
\includegraphics[width=10.0cm, angle=-90]{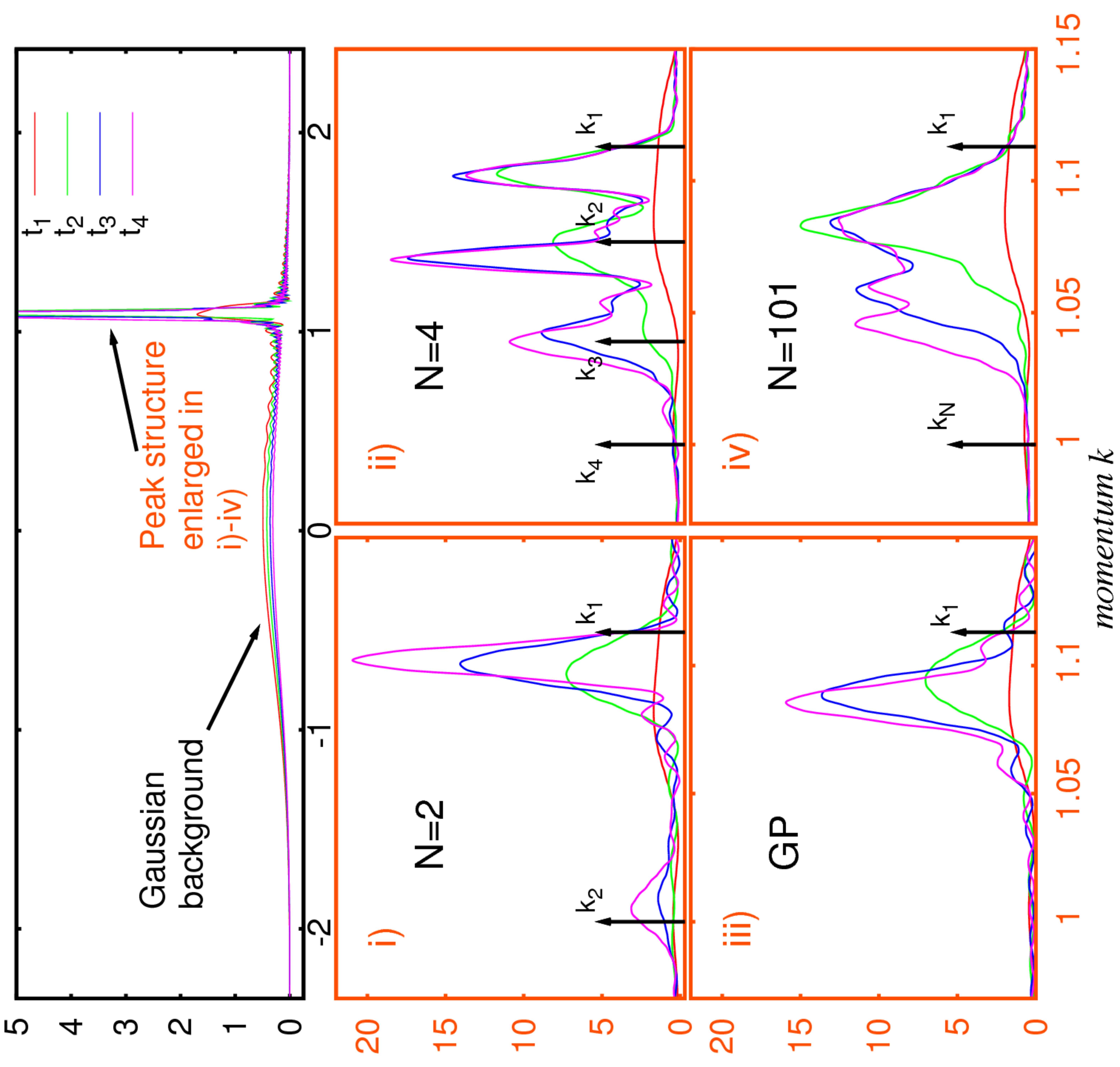}
\end{center}
\caption{{\bf The peak structures in the momentum distributions characterize the physics of many-body tunneling to open space.}
The total momentum distributions $\rho(k,t)$ for $N=101$ (top center) and their peak structures for $N=2$, $N=4$, $N=101$, and the respective Gross-Pitaevskii solutions, at times $t_1<t_2<t_3<t_4$.
The broad Gaussian-shaped backgrounds correspond to the bosons remaining in the trap,
the sharp peaks with positive momenta can be associated with the emitted bosons.
For $N=2$ we find two peaks in panel i), for $N=4$ we find three peaks and an emerging fourth peak at longer times, in panel ii).
In panel iv) we find three \textit{washed out} peaks for $N=101$. The corresponding GP dynamics reveals only a single peak for all times in iii).
The arrows in the plots mark the momenta obtained from the model consideration. All quantities shown are dimensionless.
}
\label{FIG3}
\end{figure}

The key features of the dynamics of quantum mechanical systems manifest themselves very often in characteristic momenta.
Therefore, it is worthwhile to compute and compare
evolutions of the momentum distributions $\rho(k,t)$ of our interacting bosonic systems.
Fig.~\ref{FIG3} depicts $\rho(k,t)$ for $N=2,4,101$ bosons.
At $t=0$ all the initial real space densities have Gaussian-shaped profiles resting in the internal region
(see upper panel of Fig.~\ref{FIG1}).
Therefore, their distributions in momentum space are also Gaussian-shaped and centered around $k=0$.
With time the bosons start to tunnel out of the trap. This manifests itself 
in the appearance of a pronounced peak structure on top of the Gaussian-shaped background,
see central upper panel of Fig.~\ref{FIG3}. The peak structure is very narrow -- similar to a laser or an ionization process, the bosons seem to be emitted with a very well defined momentum.
For longer propagation times a larger fraction of bosons is emitted and more
intensity is transferred to the peak structure from the Gaussian background.
Thus, we can relate the growing peak structures in the momentum distributions to the emitted bosons 
and the Gaussian background to the bosons in the source.
We decompose each momentum distribution into a Gaussian background and a peak structure to check the above relation (see Methods for details).
The integrals over the Gaussian momentum background, $P^k_{not,\rho}(t)$, 
are depicted in Fig.~\ref{FIG2} as a function of time.
The close similarity of the  $P^x_{not,\rho}(t)$, characterizing the amount of particles remaining in the internal region in real space,
and $P^k_{not,\rho}(t)$ confirms our association of the Gaussian-shaped background with source bosons and
the peak structure in the momentum distribution with emitted ones (cf. top panel of Fig.~\ref{FIG3}).

Now we are equipped to look into the mechanism of the many-body physics in the tunneling process with a simplistic model.
When the first boson is emitted to open space 
one can estimate its total energy
as an energy difference, $E^{N}-E^{N-1}$, of source systems made of $N$ and $N-1$ particles. This corresponds to assuming the negligibility of the coupling of the trapped and emitted bosons: 
$$
E_{OUT}=E^{N}-E^{N-1}\equiv \mu_1.
\nonumber
$$
This is simply the chemical potential of the parabolically trapped source system. For noninteracting particles its value would be independent of the number of bosons inside the well.
Since the boson tunnels to open space where the trapping potential is zero, it can be considered for the time being as a free particle with all its available energy, $\mu_1$, converted to kinetic energy
$E^{kin}_{OUT}=\frac{k^2}{2m}$. 
We have to expect the first emitted boson to have the momentum $k_1=\sqrt{2m\mu_1}$.
The value of the estimated momentum agrees excellently with the position of the peak in the computed exact momentum distributions,
see the arrows marked $k_1$ in Fig.~\ref{FIG3} i)-iv).
This agreement allows us to interpret the peak structures in $\rho(k,t)$ as the momenta of the emitted bosons.
As a striking feature, also other peaks with smaller $k$ appear in these spectra at later tunneling times [see Fig.~\ref{FIG3}i), ii) and iv)].
We can use an analogous argumentation and associate the second peak with the emission of the second boson and its chemical potential $\mu_2$. Its kinetic energy and the corresponding momentum, $k_2$,
can be estimated as the energy difference $\mu_2=E^{N-1} - E^{N-2}$ between the source subsystems made of $N-1$ and $N-2$ bosons, under the assumption of zero
interaction between the emitted particles and the source.
The correctness of the applied logic can be verified 
from Fig.~\ref{FIG3} i) and ii), where for the $N=2$ ($N=4$) particles the positions of the estimated momenta $k_2$ ($k_3$, $k_4$) of the second (third, fourth) emitted boson fit well with the position of the second (third, fourth) peak in the computed spectra. The momenta and details on the calculation are collected in the Supplementary Information. 
The momentum spectrum for $N=101$ bosons shows a similar behavior -- the multi-peak structures gradually develop with time
starting from a single-peak to two-peaks and so on, see Fig.~\ref{FIG3} iv).
However, from this figure we see that the positions of the peaks' maxima, and with them the momenta of the emitted bosons change with time.
On the one hand we see that the considered tunneling bosonic systems can not be utilized as an atomic laser: the intially coherent bosonic source emits particles with different, weakly time-dependent momenta. In optics such a source would be called polychromatic. On the other hand we can associate the peaks with different channels of an ionization process and their time-dependency with the channels' coupling. Thus, we conclude that it is possible to model and investigate ionization processes with tunneling ultracold bosonic systems.

Let us now investigate the coherence of the tunneling process itself.
In the above analysis of the momentum spectra we relied on the exact numerical solutions
of the TDSE for $N=2,4,101$ bosons.
We remind the reader that in the context of ultracold atoms the Gross-Pitaevskii (GP) theory
is a popular and widely used mean-field approximation describing systems under the assumption that they stay fully coherent for all times.
In our case the GP approximation assumes that the ultracold atomic cloud coherently emits the bosons to open space and keeps the source and emitted bosons coherent all the time.
To learn about the coherence properties of the ongoing dynamics it is instructive to compare exact many-body solutions of the TDSE with the idealized GP results, see Fig.~\ref{FIG3}iii).
The strengths of the inter-boson repulsion have deliberately been chosen such that the GP gives identical dynamics for all $N$ studied. 
It is clearly seen that for short initial propagation times the dynamics is indeed coherent.
The respective momentum spectra obtained at the many-body and GP levels are very similar, see Fig.~\ref{FIG3}i)-iv) for $\rho(k,t_1=100)$.
At longer propagation times ($t>t_1$), however, the spectra become considerably different. This means that with time,
the process of emission of bosons becomes less coherent.

\begin{figure}[]
\vglue -2.0 truecm
\begin{center}\hspace{1.5cm}
\includegraphics[width=5.0cm, angle=-90]{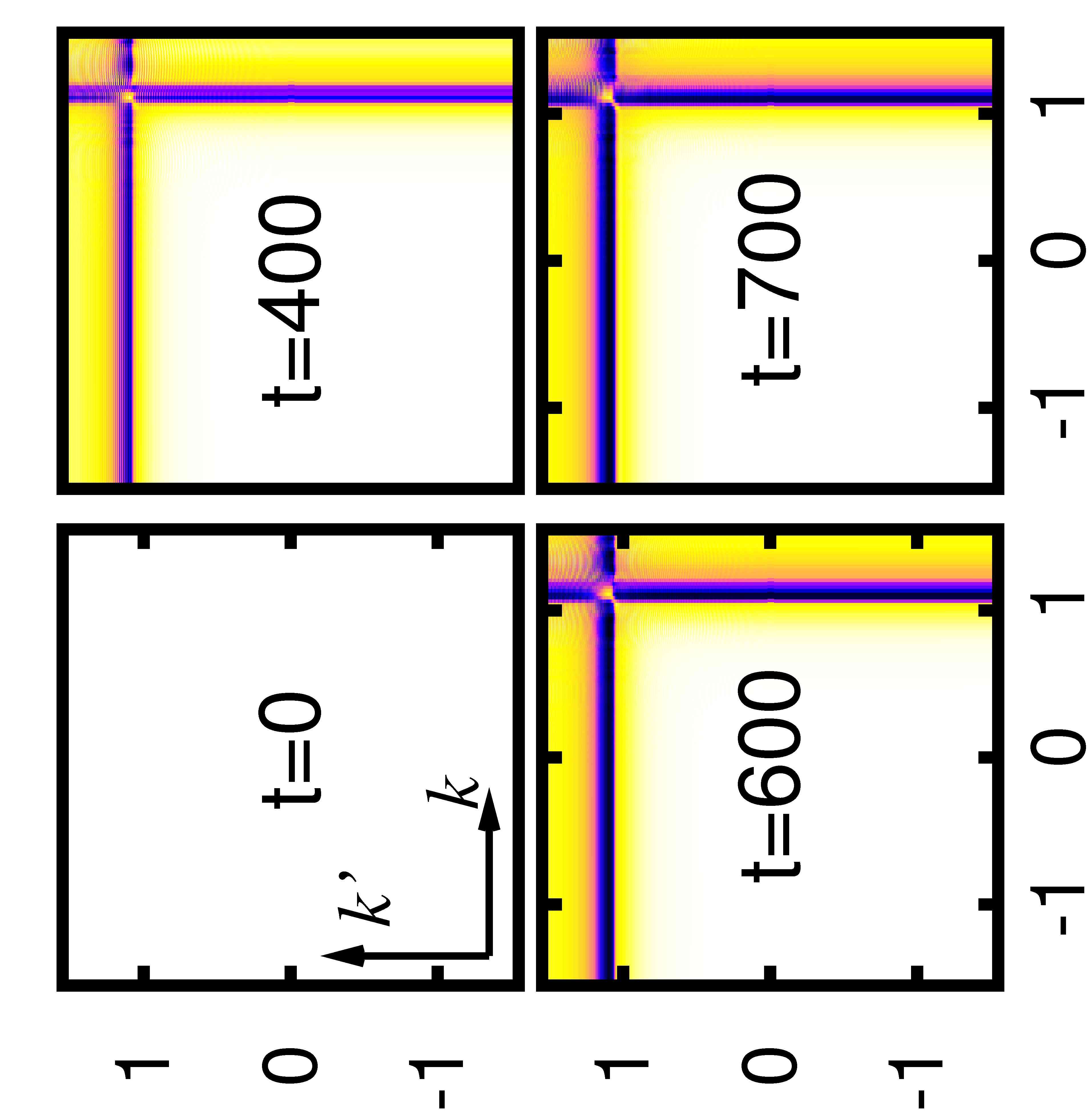}
\includegraphics[width=5.0cm, angle=-90]{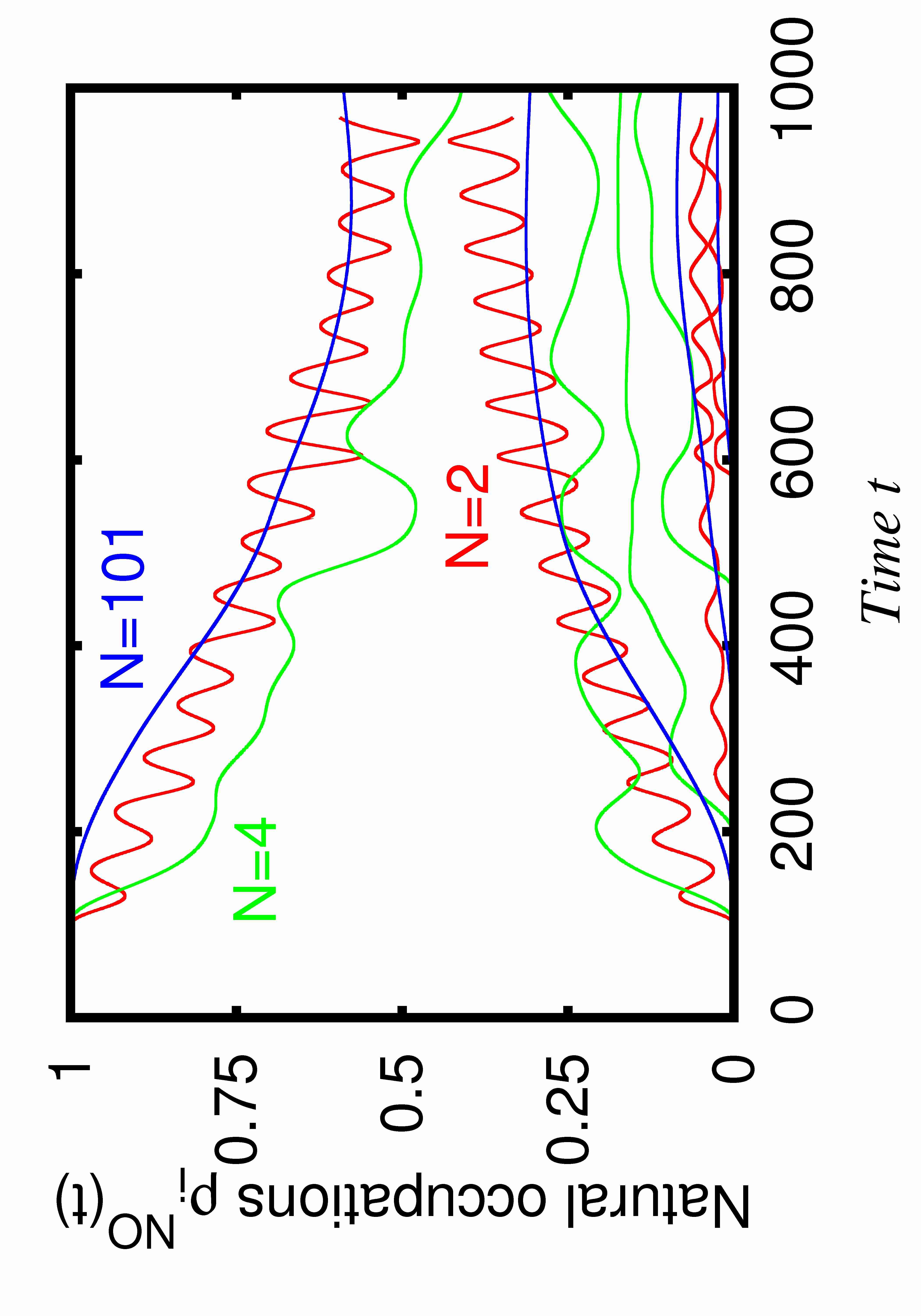}
\end{center}
\caption{{\bf Monitoring the coherence of the system.}
\textbf{Left:} The first order correlation functions in momentum space $\vert g^{(1)}(k'\vert k;t)\vert^2$ for $N=101$ 
are plotted at $t=0,400,600,700$. At $t=0$ the system is totally coherent, i.e., $\vert g^{(1)} \vert^2 =1$.
At times $t>0$, the system remains coherent everywhere in $k$-space apart from the region around $k=1$,
where we find peaks in the momentum distributions.
The loss of coherence, $\vert g^{(1)} \vert^2 \approx 0$ only in these regions allows us to conclude that the source (trapped) bosons remain coherent at all times
while the emitted ones are incoherent.
\textbf{Right:} The time evolution of the first few natural occupation numbers $\rho_{i}^{NO}(t)$ for $N=2$, $N=4$, and $N=101$ bosons.
The coherence in the systems is gradually lost with time. The systems fragment because more and more natural orbitals 
become populated. All quantities shown are dimensionless.
}
\label{FIG4}
\end{figure}

Next, to quantify the coherence and correlations between the source and emitted bosons
we compute and plot the momentum correlation functions $\vert g^{(1)}(k', k \vert t) \vert^2$
in the left panel of Fig.~\ref{FIG4} for $N=101$ (for $N=2,4$ they look almost the same). 
Let us stress here that the proper correlation properties cannot be accounted for by approximate methods. 
For example the GP solution of the problem gives $\vert g^{(1)} \vert^2 = 1$, i.e., full coherence for all times. 
For the exact solution we also obtain that at $t=0$  the system is fully coherent, and thus $\vert g^{(1)}(k' \vert k; t=0 )\vert^2 = 1$. 
Hence, the top left panel of Fig.~\ref{FIG4} is also a plot for the GP time-evolution. 
However, during the tunneling process the many-body evolution of the system becomes incoherent, i.e., $\vert g^{(1)} \vert^2 \rightarrow 0$. 
The coherence is lost \textit{only} in the momentum-space domain where the momentum distributions are peaked, the $k$-region associated with the emitted bosons (see left panel of Fig.~\ref{FIG4}).
In the remainder of $k$-space the wavefunction stays coherent for all times.
We conclude that the trapped bosons within the source remain coherent. The emitted bosons become incoherent with their source \textit{and} among each other. Therefore, the coherence between the source and the emitted bosons is lost.
A complementary argumentation with the normalized real-space correlation 
functions is deferred to the Supplementary Information and Fig.~S2 therein.

In the spirit of the seminal work of Penrose and Onsager on reduced density matrices \cite{Penrose}, we tackle the question
``How strong is the loss of coherence in many-body systems?''. The natural occupation numbers,  $\rho^{NO}_i(t)$, obtained by diagonalizing the reduced one-body density (see Methods for details) define how much the system can be described by a single, two or more quantum mechanical one-particle states. The system is condensed and coherent when only one natural occupation is macroscopic and it is fragmented when several of the $\rho^{NO}_i(t)$ are macroscopic.
In the right panel of Fig.~\ref{FIG4} we plot the evolution of the natural occupation numbers for the studied systems as a function 
of propagation time. The initial system is totally coherent -- all the bosons 
reside in one natural orbital. However, when some fraction of the bosons is emitted, a second natural orbital gradually becomes occupied. 
The decaying systems loose their coherence and become two-fold fragmented. 
For longer propagation times more natural orbitals start to be populated, indicating
that the decaying systems become even more fragmented, i.e., less coherent.
In the few-boson cases, $N=2$ and $N=4$, one can observe several stages of the development of fragmentation \cite{frag1,frag2} --
beginning with a single condensate and evolving towards the limiting form of $N$ entangled fragments in the end which means a full ``fermionization'' of the emitted particles.
In this fermionization-like case each particle will propagate with its own momentum.
For larger $N$ the number of fragments also increases with tunneling time. The details
of the evolution depend on the strength of the interparticle interaction and number of particles. By resolving the peak structure in the momentum spectra of the tunneling systems at different times
we can directly detect and quantify the evolution of coherence, correlations and fragmentation.

Finally, we tackle the intricate question whether the bosons are emitted one-by-one or several-at-a-time? 
By comparing the momentum spectra $\rho(k,t)$ depicted in Fig.~\ref{FIG3}
at different times it becomes evident that the respective peaks appear in the spectra
sequentially with time, starting from the most energetical one.
If multi-boson (two-or-more-boson) tunneling processes
would participate in the dynamics, they would give spectral features with higher momenta which are not observed in the computed spectra depicted in Fig.~\ref{FIG3}. A detailed discussion and a model are given in the Supplementary Information.
This model suggests that the bosons tunnel out {\it one-by-one}.
However, the fact that the peaks' heights and positions evolve with time, indicates that
the individual tunneling processes interfere, i.e., they are not independent. The origin behind this interference
is the interaction between the bosons.
We conclude that the overall decay by tunneling process is of a many-body nature and is formed by the interference of different single-particle tunneling processes taking place \textit{simultaneously}.

We arrive at the following physical picture of the tunneling to open space of an interacting, initially-coherent bosonic cloud.
The emission from the bosonic source is a continuous, polychromatic many-body process accompanied by a loss of coherence, i.e., fragmentation.
The dynamics can be considered as a superposition of individual single-particle tunneling processes
of a source systems with different particle numbers.
On the one hand ultracold weakly interacting bosonic clouds tunneling to open space can serve as an atomic laser, i.e., emit bosons coherently, but only for a short time. 
For longer tunneling times the emitted particles become incoherent. They lose their coherence with the source \textit{and} among each other.
On the other hand, we have shown the usage of tunneling ultracold atoms to study the dynamics of ionization processes. 
Each peak in the momentum spectrum is associated with the single particle decay of a bosonic source made of $N$, $N-1$, $N-2$, etc. particles -- in close analogy to \textit{sequential single ionization processes}.
These $N$ discrete momenta comprise a total spectrum -- in close analogy to \textit{total ionization spectra}.

As an experimental protocol for a straightforward detection of the kinetic energy of the emitted particles
one can use the presently available single atom detection techniques on atom chips \cite{single}
or
the idea of mass-spectrometry (see discussion in the Supplementary Information).
Summarizing, in many-particle systems decaying by tunneling to open space
the correlation dynamics between the source and emitted parts lead to clearly observable spectral features which are of great physical relevance.

\section*{Methods}
\subsection*{Hamiltonian and Units}
The one-dimensional $N$-boson Hamiltonian reads as follows:
$$
\hat H(x_1,\ldots,x_N) =
\sum_{j=1}^N \left(-\frac{1}{2}\frac{\partial^2}{\partial x_j^2} + V(x_j)\right) +
\sum_{j < k}^N \lambda_0 \delta(x_j-x_k).
\nonumber
$$
Here $\lambda_0>0$ is the repulsive inter-particle interaction strength
proportional to the $s$-wave scattering length $a_s$ of the bosons and $x_i$ is the coordinate of the $i$-th boson. Throughout this work $\lambda=\lambda_0(N-1)=0.3$ for all considered $N$ is used.
For convenience we work with the dimensionless quantities defined by dividing the dimensional Hamiltonian
by $\frac{\hbar^2}{m L^2}$,
where $\hbar = 1.05457 \cdot 10^{-34} \frac{\mathrm{m}^2 \mathrm{kg}}{\mathrm{sec}}$ is Planck's constant, 
$m$ is the mass of a boson and $L$ is a chosen length scale.
At $t<0$ trap is parabolic $V(x,t<0)=\frac{1}{2}x^2$,
the analytic form of $V(x,t)$ after the opening is given in ref.~\cite{No1}.

In this work we consider $^{87}$Rb atoms
for which $m = 1.44316 \cdot 10^{-25}$ kg and $a_s = 90.4a_0$ without tuning by a Feshbach resonance, 
where $a_0 = 0.0529 \cdot 10^{-9}$ m is Bohr's radius.
We emulate a quasi-1D cigar-shaped trap
in which the transverse confinement is $w_\perp = 2291.25$ Hz,
which is amenable to current experimental setups.
Following \cite{olsh}, the transverse confinement renormalizes the interaction strength.
Combining all the above,
the length scale is given by 
$L = \frac{\hbar |\lambda_0|}{2 m \omega_\perp a_s} = 1.0 \cdot 10^{-6}$ m,
and the time scale by $\frac{m L^2}{\hbar} = 1.37 \cdot 10^{-3}$ sec.

\subsubsection*{The Multiconfigurational Time-Dependent Hartree for Bosons Method}

The time-dependent many-boson wavefunction $\Psi(t)$ solving the many-boson Schr\"odinger equation $i \frac{\partial \Psi(t)}{\partial t} = \hat H \Psi(t)$
is obtained by the multiconfigurational time-dependent Hartree method 
for bosons (MCTDHB), see refs.~\cite{MCTDHB0,MCTDHB1}. Applications include new intriguing 
many-boson physics such as the death of attractive soliton trains \cite{SOL},
formation of fragmented many-body states \cite{CAT} and {\it numerically exact} double well dynamics \cite{JJ_K,Grond1,Grond2}.
Recent optimizations of the MCTDHB, see e.g. \cite{MCTDHB2,MCTDHB3}, allow now for the application of the algorithm to open systems with very large grids (here $2^{16}=65536$ basis functions), 
a particle number of up to $N=101$ and an arbitrary number of natural orbitals (here up to $8$). 
We would like to stress that even nowadays such kind of time-dependent computations are very challenging.

The mean-field wavefunction is obtained by solving
the time-dependent Gross-Pitaevskii equation, 
which is contained as a special single-orbital case in the MCTDHB equations of motion, see ref.~\cite{MCTDHB0,MCTDHB1}. 
To ensure that the tunneling wavepackets \textit{do not} reach the box borders for all presented propagation times
the simulations were done in a box $[-5, 7465]$. 
In the dimensional units we thus solve a quantum mechanical problem numerically exactly in a spatial domain extending over $8.29$mm (!).

\subsubsection*{Many-Body Analysis of the Wavefunction}

With the many-boson wavefunction $\Psi(t)$ at hand the various
quantities of interest are computed and utilized to analyze
the evolution in time of the Bose system.
The reduced one-body density matrix of the system is 
given by $\rho^{(1)}(x|x';t) = 
\langle\Psi(t)|\hat{\mathbf \Psi}^\dag(x')\hat{\mathbf \Psi}(x)|\Psi(t)\rangle$,
where $\hat{\mathbf \Psi}^\dag(x)$ is the usual
bosonic field operator creating a boson at position $x$.
Diagonalizing $\rho^{(1)}(x|x';t)$ one gets
the natural orbitals (eigenfunctions), $\phi^{NO}_i$, and
natural occupation numbers $\rho^{NO}_i$ (eigenvalues) 
from the expression $\rho^{(1)}(x|x';t)= \sum_{i=1}^M \rho^{NO}_i(t) \left(\phi^{NO}_i(x',t)\right)^*\phi^{NO}_i(x,t)$.
The latter determine the extent to which the system 
is condensed (one macroscopic eigenvalue) 
or fragmented (two or more macroscopic eigenvalues) \cite{frag1,frag2,nim}.
The diagonal part of the reduced one-body density matrix
$\rho(x,t) \equiv \rho^{(1)}(x|x';t)$ is the system's density.
The first-order correlation function
in coordinate space $g^{(1)}(x',x;t) \equiv \frac{\rho^{(1)}(x|x';t)}{\sqrt{{\rho(x,t) }{\rho(x',t)}}}$
quantifies the degree of spatial coherence of the interacting system \cite{Glauber,RDM_K}.
The respective quantities in momentum space,
such as the momentum distribution $\rho(k,t)$ and the first-order 
correlation function in momentum space
$g^{(1)}(k'\vert k;t)\equiv \frac{\rho^{(1)}(k|k';t)}{\sqrt{{\rho(k,t) }{\rho(k',t)}}}$, 
are derived from $\rho^{(1)}(x|x';t)$ via 
an application of a Fourier transform on its eigenfunctions. 

In real space the density-related nonescape probability is given by $P^x_{not,\rho}(t)=\int_{IN} \rho(x,t) dx$ (see \cite{No1} for the non-hermitian results). The \textit{momentum}-density related nonescape probability $P^k_{not,\rho}(t)$ is obtained by least-squares fitting a Gaussian function $\rho^{Gauss}(k,t)=Ae^{-(Bx)^2}$ to $\rho(k,t)$ in the $k$-space domain
$\left[-\infty,0 \right]$. $A$ and $B$ are the fit parameters.
We then define the \textit{momentum}-density related nonescape probability as
$$
 P^k_{not,\rho}(t) = \int \rho^{Gauss}(k,t) dk.
\nonumber
$$

\subsection*{Acknowledgment}

We are grateful to Marios Tsatsos for fruitful discussions. We thank Shachar Klaiman and Julian Grond for a careful reading of the manuscript and comments. 
Computation time on the bwGRiD and financial support by the HGS MathComp and the DFG also within the framework of the ``Enable fund'' of the excellence initiative at Heidelberg university are greatly acknowledged.

\newpage
\thispagestyle{empty}

\newpage
\clearpage

\begin{center}
\Large{\bf Supplementary Information} 
\end{center}

\subsection*{The Many-Body Physics of Tunneling to Open Space}

Let us commence this Supplementary Information by a model of the physical processes constituting the many-boson tunneling of an initially-coherent bosonic systems to open space. The processes are schematically depicted in Fig.~1 of the main manuscript.

\subsubsection*{Static picture: Basic processes assembling the many-body physics}
The initial and final physical situations in the ``IN'' and ``OUT'' subspaces are intuitively clear.
The totally condensed initial state lives in the ``IN'' region and is confined by a harmonic potential.
Therefore, it can be described by a harmonic oscillator-like product state, cf. e.g. ref.~[S1].
In the final state all the bosons have tunneled out and live entirely in the semi-infinite ``OUT'' region.
According to Gaudin, M (1971) {\it Phys. Rev. A} 4:386-394,
the static many-body solution of the one-dimensional bosonic system with short range repulsive interaction
on a semi-infinite axis can be constructed as a linear combination of many correlated (incoherent) states.
This implies that the dynamical final state of our system is incoherent.

To model the steps translating the fully coherent systems to complete incoherence, let us first consider the situation in which exactly one boson has tunneled through the barrier from ``IN'' to ``OUT'' and has no more connection with the interior.
The ``IN''-system now has $N-1$ particles and the ``OUT''-system has $1$ particle.
By assuming that no excitations have been produced in the ``IN''-system, 
the trapped bosons' energy is exactly reduced by the chemical potential
$\mu_1=E^N-E^{N-1},$
see bottom part of Fig.~1 of the main text.
Here $E^i$ is the energy of the trapped harmonic oscillator product state
with the distribution of $i$ bosons in the ``IN'' subspace.
We assume that the chemical potential does not depend on the number of emitted bosons, because in ``OUT'' $V(x)\approx0$.
Let us further ignore the inter-particle interaction in the exterior system.
Energy conservation requires then, that the chemical potential $\mu_1$ of the
first boson tunneled from ``IN'' to ``OUT'' region must be converted to kinetic energy.
A free particle has the kinetic energy $E^{kin}_{OUT}=\frac{k^2}{2m}$ - we thus expect the first emitted boson to have the momentum $k_1=\sqrt{2m E^{kin}_{OUT,1}}=\sqrt{2m\mu_1}$.
The above considerations imply that the many-body wavefunction can be considered in a localized basis $\vert IN ; OUT \rangle$.
The process of emission of the first boson in this basis reads
$\vert N ; 0 \rangle \rightarrow \vert N-1 ; 1^{k_1} \rangle$.
Here the $k_1$ superscript indicates that the emitted boson occupies a state which is very similar 
to a plane wave with momentum $k_1$ in the ``OUT'' subspace.
Now we can prescribe the process of the emission of the second boson as 
$\vert N-1 ; 1^{k_1} \rangle \rightarrow \vert N-2 ; 1^{k_1},1^{k_2} \rangle$.
By neglecting the interactions between the first and second emitted bosons
we can define the second chemical potential as $\mu_2=E_{IN}(N-1)-E_{IN}(N-2)$.
Thus, a second kinetic energy $E^{kin}_{OUT,2}=\mu_2$ gives rise to the momentum peak at $k_2=\sqrt{2m\mu_2}$.
Generally, the chemical potentials of the systems made of $N-i$ and $N-i-1$, $i=0,...,N-1$, particles are different,
so the corresponding peaks should appear at different positions in the momentum spectra.
We can continue to apply the above scheme until the last boson is emitted
$\vert 1 ; 1^{k_1}\cdots 1^{k_{N-1}} \rangle \rightarrow \vert 0 ; 1^{k_1} \cdots 1^{k_N}\rangle$.
The bottom panel of Fig.~1 in the main text indicates the chemical potentials for these one-particle mean-field processes by horizontal lines and the processes by the vertical arrows.
This simplified mean-field picture of the tunneling dynamics is analogous to $N$ sequential processes of ionization, where
the energetics of each independent process (channel) are defined by the chemical potential of the respective sources made of
$N$,$N-1$,$N-2$, etc. particles.

\subsubsection*{Connection of the Model to the Numerical Experiment}
Let us first compute the momenta available in the system of $N=2$ bosons with interparticle interaction strength $\lambda_0=0.3$, following ref.~[S2].
The difference between the total energies of the trapped system made of $N=2$ and $N=1$ bosons provides $k_1=1.106$.
The second momentum associated with the emition of the last boson from the parabolic trap gives $k_2=1.000$.
A similar analysis done for the system of $N=4$ bosons with the interparticle interaction strength $\lambda_0=0.1$ [$\lambda_0(N-1)=0.3$]
gives $k_1=1.106$, $k_2=1.075$, $k_3=1.038$ and $k_4=1.000$ for the first, second, third and fourth momentum respectively.
To relate the model and the full many-body results 
we draw the momenta estimated from the respective chemical potentials
in Fig.~3 of the the main text by vertical arrows.
The agreement between the momenta obtained from the model
and the respective ones from the dynamics is very good, see the arrows and the peaks in the orange framed plots in Figs.~3i) and ii).
From this figure it is clearly seen that the later in time we look at the momentum distributions $\rho(k,t)$,
the closer the peaks maxima locate to the estimated results.
Moreover, our model explains why for $N=101$ the peaks are wash out.
The chemical potentials of neighboring systems made of a big particle number ($101$ and $100$)
become very close and, as a result, the corresponding peaks start to overlap and become blurred.
Nevertheless, they are always enclosed by the first and last chemical potentials contributing, see the labels $k_1$ and $k_N$ in Fig.~3iv).

The good agreement between our model and full numerical experiments
validates the applicability of the emerged physical picture to the tunneling to open space.
We continue by excluding the possibility that the observed peaks in the momentum spectra 
can be associated with excitations inside the initial parabolic trap potential.
This can be done straightforwardly by calculating the chemical potentials associated with the configurations where 
one or several bosons reside in the second, third, etc. excited orbitals of the trapped system.
It is easy to demonstrate that the bosons emitted from these excited orbitals
would have higher kinetic energies resulting in the spectral features with higher momenta.
Since the computed spectra depicted in Fig.~3 of the main text do not reveal such spectral features
we conclude that the excitations inside the initial parabolic trap potential do not contribute to the tunneling process in a visible manner.

The above analysis suggests that the overall many-body tunneling to open space process is assembled by the elementary mean-field-like tunneling processes
analogous to the ionization of the systems made of different particle numbers which are happening \textit{simultaneously}.
We also are in the position to deduce now that every elementary contributing process is of a single-particle type.
Indeed, if it were a two-particle process, the kinetic energy of the emitted bosons 
would have been $E^{kin}_{2b}=\frac{(k^{2b}_1)^2+(k^{2b}_2)^2}{2\cdot 2m}$.
For large $N$ one can assume that the chemical potentials of the first two processes are almost equal, i.e., $\mu^{2b}_1 \approx \mu^{2b}_2 \approx 2 \mu_1$. The momentum associated with a two-particle tunneling process would be $k^{2b}_{tot}=\sqrt{4m\mu_1}=\sqrt{2} k_1$ 
-- which is far out of the domain where the peaks occur in the exact solutions.

\section*{Tracing the Coherence in Real Space}
Here we complement our study of the coherence in momentum space given in the manuscript
by its real space counterpart.
To characterize the coherence of the tunneling many-boson system in real space
we compute the normalized real space first order correlation function $g^{(1)}(x'_1\vert x_1; t)$ 
at various times $t$ for the system of $N=101$ bosons and depict the results in Fig.~S1.
From this figure we see that initially the system is fully coherent, namely $\vert g^{(1)}(x'_1\vert x_1; t=0)\vert^2=1$.
For $t>0$ $\vert g^{(1)}(x'_1\vert x_1; t)\vert^2<1$ \textit{only} in the ``OUT'' region, indicating
that only the emitted bosons quickly lose their coherence.
In contrast, the source bosons living in the interior around $x_1=x'_1=0$ remain coherent for all times.
This corroborates our findings from the first order normalized momentum correlation function
$g^{(1)}(k'_1\vert k_1; t)$ analyzed in the main text: the bosons are ejected incoherently from a source, which preserves its initial coherence.

\section*{Direct Detection of the Momentum Spectra}

It remains to line out the possible straightforward experimental verification of the emerged physical picture.
In typical experiments the bosons are ultracold many-electron atoms in a very-well defined electronic state.
According to the conjectures put forward above, the bosons will tunnel to open space with definite kinetic energy.
We propose to detect the kinetic energy of the emitted bosons 
by utilizing the techniques and principles of mass-spectrometry as schematically depicted in Fig.~S2.
One can place an ionization chamber at some distance from the trapping potential
to ionize the propagating bosonic atom suddenly.
The respective experimental ionization techniques are presently available, see e.g. Ref.~[S3] and references therein.
The now charged particle will, 
by application of a static electric field, experience a corresponding driving force and change its trajectory.
The trajectory of the ionized atom or, alternatively, the trajectory of the ionized electron
are completely described by the respective driving force, the electronic state of the atom and its initial kinetic energy.
By using a detector capable to detect the charged atom or a photoelectron multiplier for the electrons
one can monitor the deflection of the ionized particle from the initial direction of propagation.
The kinetic energy and, therefore, the momentum of the emitted boson can be calculated.
By this one can detect \textit{in situ} the momentum spectra $\rho(k,t)$ corresponding to different tunneling times and study the tunneling to open space as a function of time.

For the few-particle case it is especially interesting not only to obtain the momentum spectra, but also to monitor
the time-ordering in which the peaks appear, i.e., to monitor the time evolution of the momentum peak densities $\rho(k,t)$.
In such an experiment one can see whether the signals corresponding to the different $k_i$, $i=1,...,N$, will
be detected sequentially, starting from the largest momentum, or they appear to some degree arbitrarily.
The latter case is a clear indication that the tunneling is a combination of several single particle tunneling processes
happening {\it simultenously}, as we predict. Additionally, this measurement would be among the first direct observations 
of the dynamics of the coherence and normalized correlations in ultracold bosonic systems.

Let us summarize. The deterministic preparation of few particle ultracold systems is now possible, see Ref.~[S4].
Mass-spectrometry is one of the most well-studied techniques and working tools available and even more sophisticated detection schemes have been developed on atom chips (see Ref.~[S5]).
The combination of these facilities makes the detailed experimental time-dependent study of the tunneling mechanism feasible at present time.

\makeatletter
\renewcommand*{\@biblabel}[1]{[S#1]}
\makeatother

\newpage
\thispagestyle{empty}

\addtocounter{figure}{-4}

\renewcommand{\figurename}{Figure S\hglue -0.12 truecm}

\begin{figure}[]
    \centering
 \includegraphics[angle=-90,width=\textwidth]{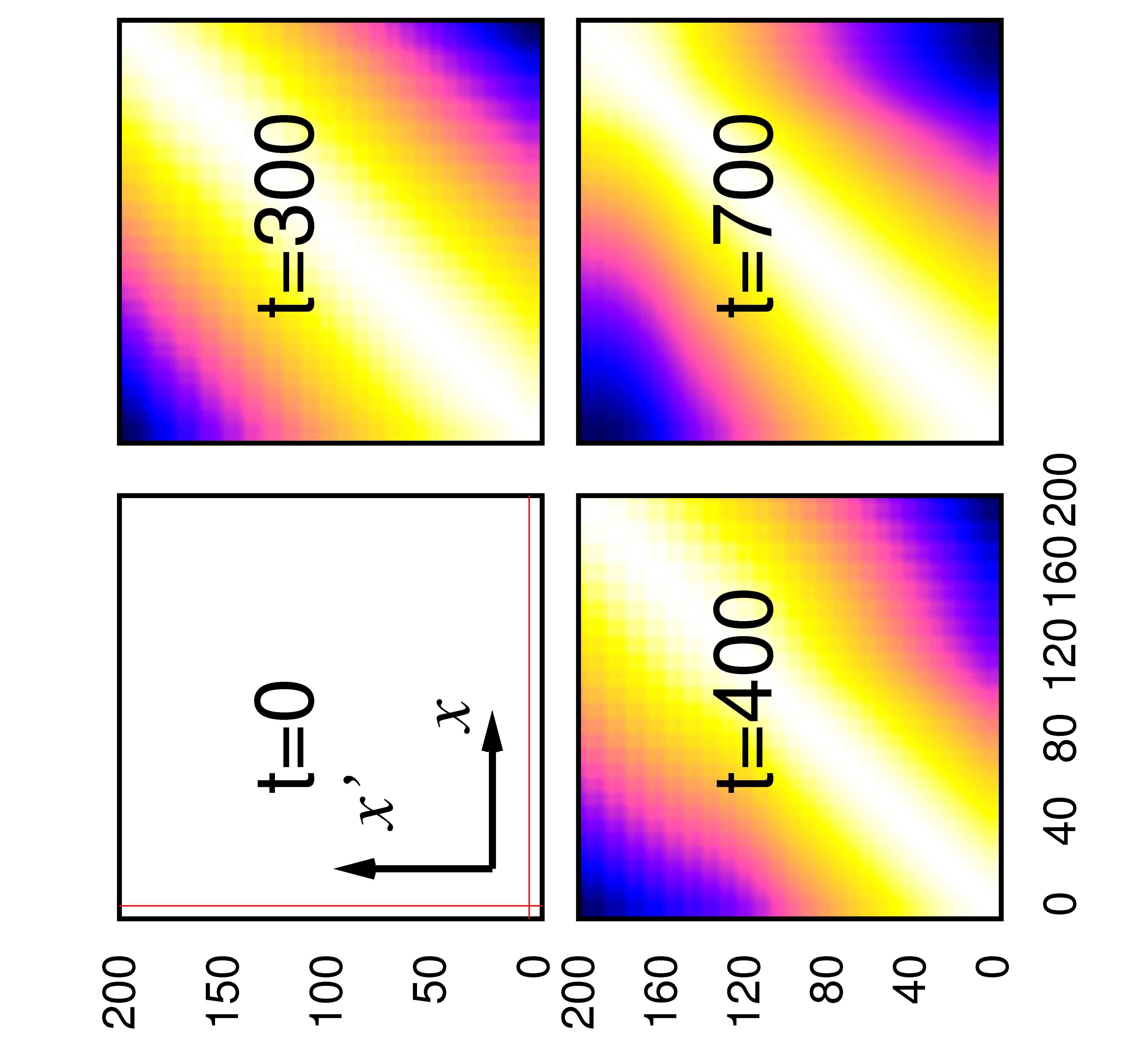} 
\caption{{\bf The real-space normalized correlation function of the tunneling to open space process:} $\vert g^{(1)}(x'_1 \vert x_1; t) \vert^2$ is used to measure the spatial coherence in the decaying system of N=101 bosons at various tunneling times. White corresponds to $\vert g^{(1)} \vert^2=1$ and black to $\vert g^{(1)} \vert^2=0$
The red lines in the top left part separate the ``IN'' and ``OUT'' regions. Here white corresponds to full coherence and black to complete incoherence. 
In the ``OUT'' region the spatial coherence is lost with time, i.e., $\vert g^{(1)} \vert^2\approx 0$ on the off-diagonal $\vert g^{(1)}(x'_1\neq x_1 \vert x_1; t) \vert^2$. 
The coherence of the source bosons is conserved, because in the ``IN'' part $\vert g^{(1)} \vert^2=1$ for all times.
See text for discussion.}
    \label{FIGS1}
\end{figure}

\begin{figure}[]
    \centering
\includegraphics[width=0.2\textwidth,angle=-90]{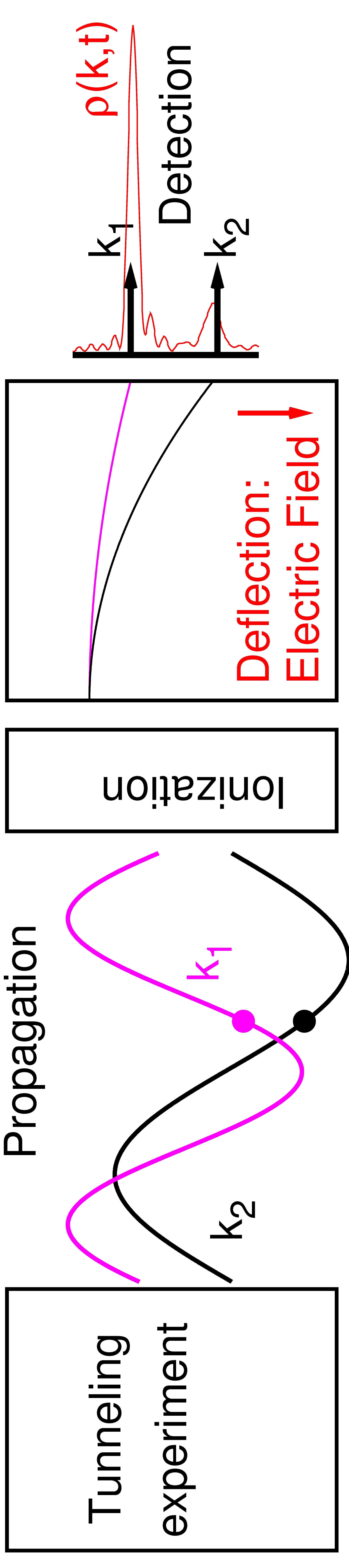}

\caption{{\bf Proposed experimental realization of the momentum spectroscopy of the many-boson system tunneling to open space.}
At some propagation distance from the experiment (left panel) the bosons are ionized by, e.g., a laser beam (center left). Subsequently, the ions/electrons are deflected by a static electric field and counted by a detector (center right). The momentum distribution can be obtained as histogram from different realizations  of the few- or many-boson tunneling process by detection of the deflected particles (right).
}
       \label{FIGS2} 
\end{figure}

\end{document}